\documentclass[twocolumn,pra,,showpacs,showkeywords,aps]{revtex4}
\usepackage[dvips]{graphicx}
\usepackage{graphicx}
\usepackage{dcolumn}
\usepackage{bm}
\newcommand{\beq}{\begin{equation}}
\newcommand{\eeq}{\end{equation}}
\newcommand{\beqa}{\begin{eqnarray}}
\newcommand{\eeqa}{\end{eqnarray}}

\begin{document}

\title{Transmission gap, Bragg-like reflection, and Goos-H\"{a}nchen shifts near the Dirac point inside a negative-zero-positive index metamaterial slab}

\author{Xi Chen$^{1,2}$\footnote{Email address: xchen@shu.edu.cn}}

\author{Li-Gang Wang$^{3,4}$\footnote{Email address: sxwlg@yahoo.com.cn}}

\author{Chun-Fang Li$^{1}$\footnote{Email address: cfli@shu.edu.cn}}

\affiliation{$^{1}$ Department of Physics, Shanghai University,
200444 Shanghai, China}

\affiliation{$^{2}$ Departamento de Qu\'{\i}mica-F\'{\i}sica,
UPV-EHU, Apdo 644, 48080 Bilbao, Spain}

\affiliation{$^{3}$ Department of Physics, The Chinese University of
Hong Kong, Shatin, New Territories, Hong Kong}

\affiliation{$^{4}$ Department of Physics, Zhejiang University,
Hangzhou 310027, China}

\begin{abstract}
Motivated by the realization of the Dirac point (DP)
with a double-cone structure for optical field in the
negative-zero-positive index metamaterial (NZPIM), the
reflection, transmission, and Goos-H\"{a}nchen (GH) shifts inside the NZPIM slab are investigated.
Due to the linear Dirac dispersion, the transmission as the function of the frequency has a gap, thus the correspond reflection has a frequency or wavelength window for the perfect reflection,
which is similar to the Bragg reflection in the one-dimensional photonic crystals.
Near the DP, the associated GH shifts in the transmission and reflection can be changed from positive to negative with increasing the wavelength. These negative and positive shifts can also be enhanced by transmission resonances, when the frequency is far from that at the DP. All these phenomena will lead to some potential applications in the integrated optics and optical devices.



\end{abstract}

\maketitle

\section{Introduction}

It is well known that a light beam totally reflected from an
interface between two dielectric media undergoes lateral shift from
the position predicted by geometrical optics \cite{Goos}. This
phenomenon was referred to as the Goos-H\"{a}nchen (GH) effect
\cite{Lotsch} and was theoretically explained firstly by Artmann in
1948 \cite{Artmann}. Up till now, the investigations of the GH
shifts have been extended to frustrated total internal reflection (FTIR)
\cite{Ghatak,Haibel,Chen-PRA}, attenuated total reflection (ATR)
\cite{Yin,Pillon}, partial reflection \cite{Hsue-T,Riesz,Li-2,Nimtz},
and other areas of physics \cite{Lotsch}, such as
quantum mechanics \cite{Renard}, acoustics \cite{Briers}, neutron
physics \cite{Ignatovich}, spintronics \cite{Chen}, atom optics
\cite{Zhang-WP} and graphene \cite{Beenakker-PRL}.

Graphene has become a subject of intense interest
\cite{Neto-GPN,Beenakker} since the graphitic sheet of one-atom
thickness has been experimentally realized by A. K. Geim \textit{et
al.} in 2004 \cite{Novoselov-GMJ}. The valence electron dynamics in
such a truly two-dimensional (2D) material is governed by a massless
Dirac equation. So graphene exhibits many unique electronic properties \cite{Neto-GPN},
including Klein tunneling \cite{Katsnelson-NG}.
On the other hand, the Dirac point (DP) in photonic crystals
(PCs) for the Bloch states \cite{Haldane,Peleg,Sepkhanov,Zhang} is found
from the similarity of the
photonic bands of the 2D PCs with the electronic bands of solids.
Several novel optical transport properties near the DP have been shown
in \cite{Peleg,Sepkhanov,Zhang}, such as conical diffraction
\cite{Peleg}, a ``pseudodiffusive" scaling \cite{Sepkhanov}, and the
photon's Zitterbewegung \cite{Zhang}. Very recently, Wang
\textit{et. al.} \cite{Wang-OL,Wang-EPL} realized the DP
with a double-cone structure for optical field in the
negative-zero-positive index metamaterial (NZPIM), and further the
pseudodiffusive property \cite{Wang-OL} and Zitterbewegung effect
\cite{Wang-EPL} near the DP inside such optically
homogenous media.

The main purpose of this paper is to investigate the transmission gap, Bragg-like reflection,
and GH shifts near the DP inside a NZPIM slab. Due to
the linear Dirac dispersion, the transmission has the frequency or wavelength stopping-band,
thus the corresponding reflection
has a frequency or wavelength window for perfect reflection,
which is analogous to Bragg-like reflection in monolayer graphene barrier \cite{Chen-APL}.
This so-called Bragg-like reflection in such a simple NZPIM slab is quite different
from that in the 1D PCs (for instance, a
stack of Bragg mirrors), resulting from the destructive and constructive
interferences. More interestingly, the associated GH shifts in the reflection and transmission can be changed from positive to negative with increasing the wavelength near the DP. Also these negative and positive shifts can be enhanced by transmission resonances, when the frequency is far from that at the DP.
All these phenomena will lead to some potential applications in the integrated optics and optical devices, such
as frequency or wavelength filters and frequency-dependent spatial modulator.

\section{Model}
\label{Model}

\begin{figure}[tt]
\begin{center}
\scalebox{0.88}[0.78]{\includegraphics{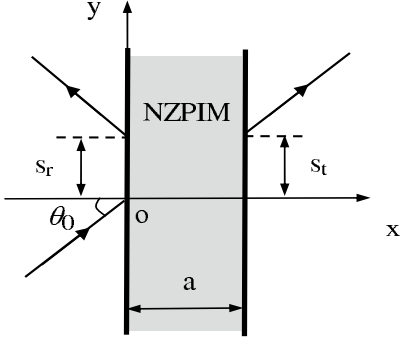}}
\end{center}
\caption{\label{configuration} Schematic diagram of the reflection and transmission
inside the NZPIM
slab configuration, where $s_r$ and $s_t$ denotes the GH shifts for the reflected and transmitted light beams, respectively.}
\end{figure}

For simplicity, we consider a TE-polarized light beam with angular frequency $\omega$ and incidence angle $\theta_0$
upon the NZPIM slab in the vacuum, as shown in Fig. \ref{configuration}, the dispersion of the NZPIM
has a linear dispersion \cite{Wang-OL}
\beq
\label{dispersion}
k(\omega)=(\omega-\omega_D)/v_D,
\eeq
with group velocity $v_D= (d \omega/dk)|_{\omega=\omega_D}$, $\omega_D$
is the frequency of the DP (corresponding wavelength is $\lambda_D = 2 \pi c/ \omega_D$), where two bands touch each other forming a double-cone structure. Near the DP,
the light transport obeys the massless Dirac equation as follows:
\beqa
\left[\begin{array}{cc}
0 & - i (\frac{\partial}{\partial x} -i\frac{\partial}{\partial y}) \\
 -i (\frac{\partial}{\partial x} + i\frac{\partial}{\partial y}) & 0
\end{array}\right] \Psi
=\left(\frac{\omega-\omega_D}{v_D}\right) \Psi,
\eeqa
where $\Psi= \left(\begin{array}{cc}
E_{z1}(x,y,\omega)  \\
E_{z2}(x,y,\omega)
\end{array}\right)$ are the eigenfunctions of the electric fields
with the same $k(\omega)$. It is noted that the condition for realization
of the DP in the homogenous optical medium is the index varying from negative to zero
and then to positive with frequency \cite{Wang-OL}, which is called as NZPIM.
For simplicity, we take the Drude model as the parameters for both the relative permittivity
and permeability of the NZPIM \cite{Wang-OL,Wang-EPL}:
\beq
\varepsilon_1({\omega})=1-\omega^2_{ep}/(\omega^2+i\gamma_e \omega),
\eeq
\beq
\mu_1({\omega})=1-\omega^2_{mp}/(\omega^2+i\gamma_m \omega),
\eeq
where $\omega^2_{ep}$ and $\omega^2_{mp}$ are the electronic and magnetic plasma frequencies,
and $\gamma_e$ and $\gamma_m$ are the damping rates relating to the absorption of the material.
Here we can assume $\gamma_e=\gamma_e=\gamma \ll \omega^2_{ep},\omega^2_{mp}$. It is important
that when $\omega_{ep}=\omega_{mp}=\omega_D$ and $\gamma =0$ (no loss), then both $\varepsilon_1(\omega_D)$
and $\mu_1(\omega_D)$ may be zero simultaneously.
In this case, we find $k(w_D)\approx 0$
and $v_D\simeq c/2$, where $c$ is the light speed in vacuum \cite{Wang-EPL}.
In what follows we will discuss the reflection, transmission, and the associated GH shifts
near the DP in NZPIM slab.

\section{Reflection and Transmission}

In this section, we will firstly investigate the properties of the reflection and transmission.
Assuming the incident plane wave,
$E^{in}_{z}(x,y)= \exp{[i(k_x x +k_y y)]}$, where $k_x= k_0 \cos \theta_0$,
$k_y= k_0 \sin \theta_0$, $k_0 = \omega/c$ is the wave vector in vacuum, the reflected and transmitted plane waves can be expressed by
$E^{ref}_{z}(x,y)= r \exp{[i(- k_x x +k_y y)]}$ and $E^{tr}_{z}(x,y)= t \exp{\{i[- k_x (x-a) +k_y y]\}}$,
where the reflection coefficient
$r$ is
\begin{eqnarray}
\label{r}
r &&=\frac{\exp(i\pi/2)}{4g^2}\left(\frac{\mu_0}{\mu_1}\frac{k_{1x}}{k_x}-\frac{\mu_1}{\mu_0}\frac{
k_x}{k_{1x}}\right) \nonumber \\ &&\times\left[\sin2k_{1x} a
+i\left(\frac{\mu_1}{\mu_0}\frac{k_x}{k_{1x}}+
\frac{\mu_0}{\mu_1}\frac{k_{1x}}{k_x}\right)\sin^2 k_{1x} a\right],
\end{eqnarray}
and the transmission coefficient is
$
t= e^{i \phi}/g
$
with the following complex number,
$$
ge^{i \phi}=\cos k_{1x} a+
\frac{i}{2}\left(\frac{\mu_1}{\mu_0}\frac{k_x}{k_{1x}}+
\frac{\mu_0}{\mu_1}\frac{k_{1x}}{k_x}\right) \sin k_{1x} a,
$$
with $k_{1x}= \sqrt{k^2_1- k^2_y}$ and $k_1 = (\omega-\omega_D)/v_D$ near the DP.
It is clear that the wave vector
$k_{1x}$ depends on the different frequencies $\omega$ and parallel wave vector $k_y$,
which will resulting the unique properties of reflection and transmission in two cases of $\omega > \omega_D$
and $\omega < \omega_D$.

\textit{Case 1}: $\omega > \omega_D$. The reflection probability $R$ can be given by
Eq. (\ref{r}),
\begin{equation}
R \equiv|r|^2= \frac{1}{4g^2}\left(\frac{\mu_0}{\mu_1}\frac{k_{1x}}{k_x}-\frac{\mu_1}{\mu_0}\frac{
k_x}{k_{1x}}\right)^2 \sin^2k_{1x}a,
\end{equation}
and the transmission probability $T=1-R$ is also given by
$T \equiv |t|^2= 1/g^2$. Under resonance conditions,
$k_{1x} a = N \pi$, ($N=0, 1,...$), the reflection probability $R_{min}$ reaches
the zero and the transmission probability $T_{max}$ is equal to $1$.
Otherwise, at the anti-resonances, $k_{1x} a = (N+1/2) \pi$, ($N=0, 1,...$)
the reflection probability $R$ tends to
\beq
R_{max} = \left(\frac{\mu_0}{\mu_1}\frac{k_{1x}}{k_x}+\frac{\mu_1}{\mu_0}\frac{
k_x}{k_{1x}}\right)/\left(\frac{\mu_0}{\mu_1}\frac{k_{1x}}{k_x}-\frac{\mu_1}{\mu_0}\frac{
k_x}{k_{1x}}\right),
\eeq
and the corresponding transmission probability $T_{min}$ is equal to
\beq
T_{min} = 4/ \left(\frac{\mu_0}{\mu_1}\frac{k_{1x}}{k_x}+\frac{\mu_1}{\mu_0}\frac{
k_x}{k_{1x}}\right)^2,
\eeq
However, we emphasize here that the reflection and transmission can be divided into
evanescent and propagating modes, taking the influence of the incidence angle $\theta_0$ into account.
The propagation of the light beam inside the NZPIM slab can be evanescent when $\theta_0 > \theta_c$,
where the critical angle for total reflection can be defined as
\begin{equation}
\label{critical angle} \theta_c = \sin^{-1}
\left[2 \left(1- \frac{\omega_D}{\omega}\right)\right],
\end{equation}
with the necessary condition $\omega_D < \omega <2 \omega_D$. In this case, the
transmission and reflection probabilities damped exponentially in the following form:
\begin{equation}
\label{decay-t1}
T \approx  \frac{e^{- 2 \kappa a}}{1+ \frac{1}{4}\left(\frac{\mu_1}{\mu_0}\frac{k_x}{k_{1x}}+
\frac{\mu_0}{\mu_1}\frac{k_{1x}}{k_x}\right)},
\end{equation}
and
\begin{equation}
\label{decay-r1}
R \approx 1-  \frac{e^{- 2 \kappa a}}{1+ \frac{1}{4}\left(\frac{\mu_1}{\mu_0}\frac{k_x}{k_{1x}}+
\frac{\mu_0}{\mu_1}\frac{k_{1x}}{k_x}\right)},
\end{equation}
where $\kappa = [k^2_y - k^2_1]^{1/2}$ is the decay
constant. As a matter of fact, the light beam can transmit though the NZPIM
slab in propagating mode at any incidence angles,
when the critical angle $\theta_c$ is no longer valid for $\omega>2\omega_D$.

\textit{Case 2}: $\omega < \omega_D$. The reflection and transmission
probability can be also damped exponentially when the
incidence angle $\theta_0$ is larger than the critical
angle,
\begin{equation}
\label{critical angle-2}
\theta'_c = \sin^{-1}
\left[2\left(\frac{\omega_D}{\omega}-1\right)\right],
\end{equation}
with the necessary condition $\frac{2}{3} \omega_D < \omega <\omega_D$.
On the contrary, the reflection and transmission probabilities will
oscillate periodically on the thickness $a$ of the slab, as mentioned \textit{Case I}.
In this case, the refractive index, defined as  $n_1=-\sqrt{\varepsilon_1 \mu_1}$,
should be negative, which will lead to the negative GH shifts, as discussed later.

\begin{center}
\begin{figure}[]
\scalebox{0.32}[0.32]{\includegraphics{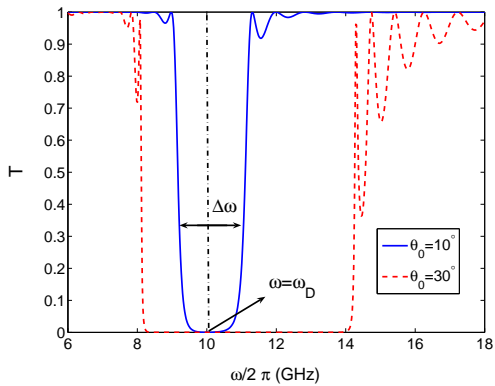}}
\caption{(Color online)\label{transmissiongap} The transmission gap as the function of the frequency $\omega$,
where $a=100$ mm, and $\omega_D=10\times 2 \pi$ GHz. Solid and dashed curves correspond to $\theta_0=10^{\circ}$ and $\theta_0=30^{\circ}$.}
\end{figure}\end{center}

Based on the mentioned-above properties of the reflection and transmission
in these two cases, the transmission as the function of frequency $\omega$
has a gap, as shown in Fig. \ref{transmissiongap}, where $a=100$ mm, and $\omega_D=10\times 2 \pi$ GHz. Solid and dashed curves correspond to $\theta_0=10^{\circ}$ and $\theta_0=30^{\circ}$.
Similarly, the transmission also have a stopping-band for the wavelength,
since $\lambda=2 \pi \omega /c$.  Since $k^2_{1x} = (\omega-\omega_D)^2/v^2_D -k^2_y <0$,
the frequency region of the
transmission gap in Fig. \ref{transmissiongap} is given by $ \omega_D- k_y v_D  < \omega < \omega_D +
k_y v_D, $ which leads to the width of transmission gap as follows,
\begin{equation}
\label{gap}
\Delta \omega = 2 k_y v_D.
\end{equation}
This means $\Delta\omega/\omega=\sin\theta_0$ with the help of $v_D\simeq c/2$.
It is further shown that the transmission gap with the center $\omega= \omega_D$ becomes
narrower with the decrease of the incidence angle, and even vanishes
at normal incidence. This transmission gap, which is analogous to that in single graphene barrier
\cite{Chen-APL}, is due to the evanescent
waves in two cases of $\omega > \omega_D$ and $\omega < \omega_D$.
Furthermore, the tunable transmission gap
can be further understood by the dependence
of the critical angle on the frequency, $\omega$.

\begin{center}
\begin{figure}[]
\scalebox{0.32}[0.32]{\includegraphics{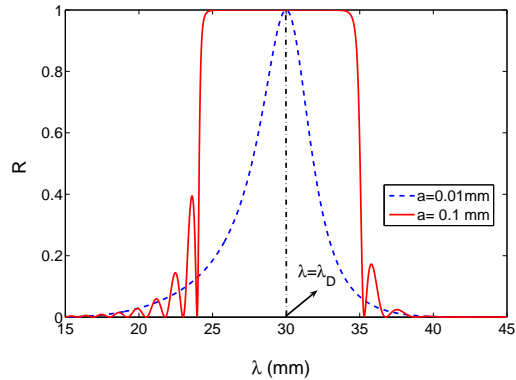}}
\caption{(Color online)\label{reflection} The reflection probability $R$ as the function of the wavelength  $\lambda$,
where $\theta_0=20^{\circ}$, and other parameters are the same as in Fig. \ref{transmissiongap}. Solid and dashed curves correspond to $a=100$ mm  and $a=10$ mm.}
\end{figure}\end{center}

Fig. \ref{reflection} indicates the dependence of
corresponding reflection probability $R$ on the wavelength $\lambda = 2 \pi\omega/c $, where $\theta_0=20^{\circ}$, and other parameters are the same as in Fig. \ref{transmissiongap}. Solid and dashed curves correspond to $a=100$ mm  and $a=10$ mm. It is interesting that the light beam can be perfectly reflected by
such single NZPIM slab at some range of the wavelength. As indicated in Fig. \ref{reflection},
the wavelength window for perfect reflection will become narrower with the increase of
the width of slab, caused by the decay factor $\exp{(-
2 \kappa a)}$ in Eqs. (\ref{decay-t1}) and (\ref{decay-r1}).
It is clearly seen from Fig. \ref{transmissiongap} that the reflection also has a similar frequency window for the
perfect reflection, since $R=1-T$.
These frequency or wavelength passing-band in reflected discussed here
is similar to but different from the Bragg reflection in the 1D PCs. This so-called Bragg-like reflection discussed here is exactly due to the linear Dirac dispersion described by Eq. (\ref{dispersion}), which results in the evanescent waves in two cases of $\omega> \omega_D$ and $\omega<\omega_D$, corresponding to the two eigenfunctions
of electric fields with the same $k(\omega)$. In a word,
the Bragg-like reflection will provide alternative way to realize the frequency
or wavelength filters with more design flexibility and miniaturization.

\section{Goos-H\"{a}nchen shifts}

\begin{center}
\begin{figure}[]
\scalebox{0.32}[0.32]{\includegraphics{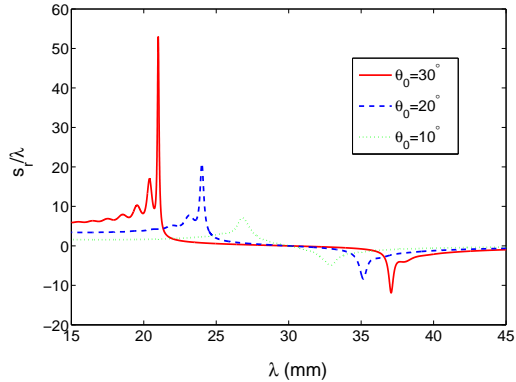}}
\caption{(Color online)\label{GHshift1} The GH shifts as the function of the wavelength, $\lambda$,
where $a=100$ mm, and other parameters are the same as in Fig. \ref{transmissiongap}.
Solid, dashed and dotted curves correspond to $\theta_0=30^{\circ}$, $\theta_0=20^{\circ}$, and $\theta_0=10^{\circ}$.}
\end{figure}
\end{center}

Now, we have a look at the GH shifts in the reflection and transmission inside the single NZPIM slab.
When a well-collimated light beam with the central incidence angle $\theta_0$ is considered, 
the GH shifts in reflection and transmission, according to Artman's stationary phase method \cite{Artmann}, can be defined as
\beq
s_{r,t}=-\frac{\partial \phi_{r,t}}{\partial k_{y}}|_{\theta=\theta_0},
\eeq
where $k_y = k_0 \sin \theta$, $\theta$ represents the incidence angle of the plane wave component under consideration, $\phi_r = \phi+\pi/2$ and $\phi_r=\phi$ are the phase shifts of the reflected and transmitted light beams, respectively.
Clearly, the GH shift in transmission is the equal to that in reflection inside such symmetric slab configuration, because the values of the derivation of the phase shifts with respect to $k_y$ are the same.
Fig. \ref{GHshift1} shows that the GH shifts can be positive and negative, where $a=100$ mm, and other parameters are the same as in Fig. \ref{transmissiongap}. Solid, dashed and dotted curves correspond to $\theta_0=30^{\circ}$, $\theta_0=20^{\circ}$, and $\theta_0=10^{\circ}$. It is shown that the GH shifts can be positive for $\lambda<\lambda_D$, while they can be negative for $\lambda > \lambda_D$. More interestingly, the GH shifts near the DP can change from positive to negative with the increase (decrease) of the wavelength (frequency).

\begin{center}
\begin{figure}[t]
\scalebox{0.32}[0.32]{\includegraphics{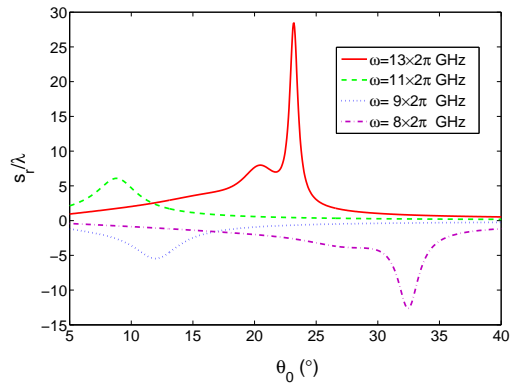}}
\caption{(Color online)\label{GHshift2} The GH shifts as the function of incidence angle $\theta_0$,
where $a=100$ mm, and other parameters are the same as in Fig. \ref{transmissiongap}.
Solid, dashed, dotted and dot-dashed curves correspond to $\omega= 13 \times 2 \pi$ GHz, $\omega= 11 \times 2 \pi$ GHz, $\omega= 9 \times 2 \pi$ GHz, and $\omega= 8 \times 2 \pi$ GHz.}
\end{figure}
\end{center}
Fig. \ref{GHshift2} also shows the dependence of the GH shifts on the incident angle $\theta_0$,
where $a=100$ mm, and other parameters are the same as in Fig. \ref{transmissiongap}.
Solid, dashed, dotted and dot-dashed curves correspond to $\omega= 13 \times 2 \pi$ GHz, $\omega= 11 \times 2 \pi$ GHz, $\omega= 9 \times 2 \pi$ GHz, and $\omega= 8 \times 2 \pi$ GHz.
It is reasonable that the GH shifts
can be negative in the case of $\omega < \omega_D$, where the refractive index $n_1= -\sqrt{\varepsilon_1 \mu_1}$
is negative, while the GH shifts are positive in the case of $\omega > \omega_D$, where the refractive index $n_1= \sqrt{\varepsilon_1 \mu_1}$
is positive. In addition, it is also shown that the GH shifts near the DP have only the order of wavelength due to the evanescent waves. The smallness of the GH shifts are similar to those in total reflection or FTIR structure.
However, when the frequency is far from that at the DP, the incidence angle will be less than
the critical angle, or there is no critical angle as discussed above. Thus,
the negative and positive GH shifts can also be enhanced by the transmission resonances, as shown in
Figs. \ref{GHshift1} and \ref{GHshift2}. In addition, the GH shifts also depends on the width $a$ of the slab. It
can be predicted from Ref. \cite{Chen-PRA} that the negative and positive GH shifts in the evanescent case will saturate to a constant with increasing the slab's width, when the incidence angle is larger than the critical angle.
In a word, these negative and positive GH shifts are applicable to realize the frequency or wavelength filters in spatial domain, and frequency-dependent spatial modulator.

\section{Conclusion}
In conclusion, we have investigated the transmission gap, Bragg-like reflection, and the associated GH shifts inside the NZPIM slab. It is found that the transmission has a frequency stopping-band,
thus the corresponding reflection
has a frequency or wavelength window for the perfect reflection.
This so-called Bragg-like reflection, resulting from the linear Dirac dispersion of NZPIM, is similar to but different from the Bragg reflection in the 1D PCs.
In addition, the GH shifts near the DP can be changed from positive to negative with increasing wavelength, based on the unique properties of the reflection and transmissions. These negative and positive shifts can also be enhanced by transmission resonances, when the frequency is far from that at the DP, $\omega_D$. With the experimental realization of the NZPIM \cite{Zhang-JAP}, we hope these phenomena will lead to some applications in the integrated optics and optical devices.

\section*{Acknowledgements}
This work is supported by the National Natural Science Foundation of
China (Grants No. 60806041, No. 10604047, and No.
60877055), the Shanghai Rising-Star Program
(Grants No. 08QA14030), the Science and Technology Commission of Shanghai
Municipal (Grants No. 08JC14097), the Shanghai Educational Development
Foundation (Grants No. 2007CG52), and the Shanghai Leading Academic Discipline
Program (Grants No. S30105). X. C. is also supported by Juan de la Cierva
Programme of Spanish MICINN. L.-G. W. would like to thank the supports from CUHK 2060360 and RGC 403609.

\end{document}